\documentclass[12pt]{article}
\usepackage{pdproc,epsfig} 
\usepackage{graphicx, multicol, color}
\usepackage{marvosym, amsfonts, floatflt,enumerate}
\begin{document}
\newcommand{\dd}{\displaystyle}
\newcommand{\om}{\omega}
\newcommand{\lbarr}{\overline{\ell}_R}
\newcommand{\lbar}{\overline{\ell}}
\newcommand{\ebarr}{\overline{e}_R}
\newcommand{\ebar}{\overline{e}}
\newcommand{\mbarr}{\overline{\mu}_R}
\newcommand{\mbar}{\overline{\mu}}
\newcommand{\tbarr}{\overline{\tau}_R}
\newcommand{\tbar}{\overline{\tau}}
\newcommand{\nubarr}{\overline{\nu}_R}
\newcommand{\nubar}{\overline{\nu}}
\newcommand{\qbar}{\overline{q}}
\newcommand{\qubarr}{\overline{u}_R}
\newcommand{\qubar}{\overline{u}}
\newcommand{\qcbarr}{\overline{c}_R}
\newcommand{\qcbar}{\overline{c}}
\newcommand{\qtbarr}{\overline{t}_R}
\newcommand{\qtbar}{\overline{t}}
\newcommand{\qdbarr}{\overline{d}_R}
\newcommand{\qdbar}{\overline{d}}
\newcommand{\qsbarr}{\overline{s}_R}
\newcommand{\qsbar}{\overline{s}}
\newcommand{\qbbarr}{\overline{b}_R}
\newcommand{\qbbar}{\overline{b}}
\newcommand{\phit}{\varphi_T}
\newcommand{\phitu}{\varphi_{T_1}}
\newcommand{\phitd}{\varphi_{T_2}}
\newcommand{\phitt}{\varphi_{T_3}}
\newcommand{\phitl}{\varphi_T^{(l)}}
\newcommand{\phitlu}{\varphi_{T_1}^{(l)}}
\newcommand{\phitld}{\varphi_{T_2}^{(l)}}
\newcommand{\phitlt}{\varphi_{T_3}^{(l)}}
\newcommand{\phitq}{\varphi_T^{(q)}}
\newcommand{\phitqu}{\varphi_{T_1}^{(q)}}
\newcommand{\phitqd}{\varphi_{T_2}^{(q)}}
\newcommand{\phitqt}{\varphi_{T_3}^{(q)}}
\newcommand{\phis}{\varphi_S}
\newcommand{\phisu}{\varphi_{S_1}}
\newcommand{\phisd}{\varphi_{S_2}}
\newcommand{\phist}{\varphi_{S_3}}
\newcommand{\phitz}{\varphi_T^0}
\newcommand{\phitzu}{\varphi^0_{T_1}}
\newcommand{\phitzd}{\varphi^0_{T_2}}
\newcommand{\phitzt}{\varphi^0_{T_3}}
\newcommand{\phisz}{\varphi_S^0}
\newcommand{\phiszu}{\varphi^0_{S_1}}
\newcommand{\phiszd}{\varphi^0_{S_2}}
\newcommand{\phiszt}{\varphi^0_{S_3}}
\newcommand{\xit}{\tilde{\xi}}
\newcommand{\xip}{\xi^\prime}
\newcommand{\xipp}{\xi^{\prime\prime}}
\newcommand{\xippt}{\tilde{\xi}^{\prime\prime}}
\newcommand{\xiz}{\xi^0}
\newcommand{\vt}{v_\text{T}}
\newcommand{\vs}{v_\text{S}}
\newcommand{\vtl}{v^{(l)}_\text{T}}
\newcommand{\vtq}{v^{(q)}_\text{T}}
\newcommand{\ep}{\varepsilon^\prime}
\newcommand{\epp}{\varepsilon^{\prime\prime}}
\newcommand{\up}{u^{\prime}}
\newcommand{\upp}{u^{\prime\prime}}
\newcommand{\etau}{\eta_1}
\newcommand{\etad}{\eta_2}
\newcommand{\yup}{y_u^{\prime}}
\newcommand{\ydp}{y_d^{\prime}}
\newcommand{\ycp}{y_c^{\prime}}
\newcommand{\ysp}{y_s^{\prime}}
\newcommand{\ytp}{y_t^{\prime}}
\newcommand{\ybp}{y_b^{\prime}}
\newcommand{\phiUdu}{\phi^\text{U}_{21}}
\newcommand{\phiUdt}{\phi^\text{U}_{23}}
\newcommand{\phiDtd}{\phi^\text{D}_{32}}
\newcommand{\phiDdt}{\phi^\text{D}_{23}}
\newcommand{\Uu}{\text{U(1)}}
\newcommand{\Ud}{\text{U(2)}}
\newcommand{\SUd}{\text{SU(2)}}
\newcommand{\SUt}{\text{SU(3)}}
\newcommand{\SUc}{\text{SU(5)}}
\newcommand{\SOt}{\text{SO(3)}}
\newcommand{\SOdie}{\text{SO(10)}}
\newcommand{\Aq}{\text{A}_4}
\newcommand{\Zd}{\text{Z}_2}
\newcommand{\Zt}{\text{Z}_3}
\newcommand{\Zs}{\text{Z}_6}
\newcommand{\Zn}{\text{Z}_9}
\newcommand{\Tp}{\text{T}^{\prime}}
\newcommand{\St}{\text{S}_3}
\newcommand{\Sq}{\text{S}_4}
\newcommand{\Dq}{\text{D}_4}
\renewcommand{\thefootnote}{\alph{footnote}}
\newcommand{\ket}[1]{|#1\rangle }
\newcommand{\bra}[1]{\langle#1| }
\newcommand{\braket}[2]{\langle#1|#2\rangle }
\newcommand{\mean}[1]{\langle#1\rangle}
\newcommand{\derp}{\partial}
\newcommand{\onep}{1^\prime}
\newcommand{\onepp}{1^{\prime\prime}}
\newcommand{\twop}{2^\prime}
\newcommand{\twopp}{2^{\prime\prime}}
\newcommand{\brap}[1]{(#1)^\prime}
\newcommand{\brapp}[1]{(#1)^{\prime\prime}}
\newcommand{\bras}[1]{(#1)_S}
\newcommand{\braa}[1]{(#1)_A}
 
\title{
 THEORY OF THE NEUTRINO MASS}

\author{FERRUCCIO FERUGLIO}

\address{Dipartimento di Fisica `G.~Galilei', Universit\`a di Padova\\
INFN, Sezione di Padova, Via Marzolo~8, I-35131 Padua, Italy\\
 {\rm E-mail: feruglio@pd.infn.it}}

\author{CLAUDIA HAGEDORN}

\address{Max-Planck-Institut f\"ur Kernphysik\\
Postfach 10 39 80, 69029 Heidelberg, Germany\\
 {\rm E-mail: hagedorn@mpi-hd.mpg.de}}


\author{YIN LIN and LUCA MERLO}

\address{Dipartimento di Fisica `G.~Galilei', Universit\`a di Padova\\
INFN, Sezione di Padova, Via Marzolo~8, I-35131 Padua, Italy\\
 {\rm E-mail: yin.lin@pd.infn.it,merlo@pd.infn.it}}

\abstract{Theoretical aspects of neutrino physics are reviewed, with emphasis on possible explanations of the smallness of neutrino masses and of the peculiar mixing pattern observed in the lepton sector.
Some theoretically motivated frameworks, such as those based on spontaneously broken discrete flavour symmetries, are discussed, stressing the importance of
low-energy observables, like anomalous magnetic moments, electric dipole moments and lepton flavour violating transitions, to test further features of these models. }
   
\normalsize\baselineskip=15pt
\section{Introduction}
Most of our current knowledge of neutrino properties comes from the study of neutrino propagation
over macroscopic distances, ranging from several hundred meters to astronomical distances. 
The experiments have exploited neutrinos coming from four independent sources:
stars, cosmic rays, reactors and particle accelerators,
thus requiring a fruitful interplay among different branches of science.
The observed anomalies with respect to the neutrino properties predicted
by the Glashow-Weinberg-Salam Standard Model (SM) have found the simplest possible interpretation:
to a very good degree of accuracy neutrinos propagate as a set of three massive neutral fermions, 
with a peculiar pattern of mixing angles between
mass and interaction eigenstates. Such a solution is supported by data
spanning more than twelve orders of magnitude in the relevant variable, $L/E$.

This important result can be encoded in a Lorentz
invariant  Lagrangian of the type
\begin{equation}
{\cal L}={\cal L}_{SM}+\delta{\cal L}(m_\nu)+...~~~,
\end{equation}
where ${\cal L}_{SM}$ is the SM Lagrangian possessing the three global, non-anomalous, conserved
quantum numbers $B/3-L_i$, ($i=e,\mu,\tau$), while $\delta{\cal L}(m_\nu)$ is an additional term that describes
neutrino masses, provides the first and up to now only evidence for physics beyond the SM and breaks the three SM global charges
with the possible exception of $B-L$ \cite{fabio}. Dots stand for a collection of additional
operators, related to other possible small effects such as
neutrino decay, decoherence, spin flavour precession, non-standard neutrino interactions, mass varying neutrinos
and so on, giving negligible contribution to neutrino propagation in current experiments.
We recall that $\delta{\cal L}(m_\nu)$ should describe the following data \cite{fogli}:
\begin{eqnarray}
&\Delta m^2_{sol}=(7.66\pm 0.35)\times 10^{-5}~ {\rm eV}^2&\nonumber\\
&\Delta m^2_{atm}=(2.38\pm 0.27)\times 10^{-3}~ {\rm eV}^2&\nonumber\\
&\sin^2\theta_{13}<0.032~~~(\theta_{13}<10.3^0)~~(2\sigma)&\nonumber\\
&\sin^2\theta_{23}=0.45^{+0.16}_{-0.09}~~~(\theta_{23}=(42.1^{+9.2}_{-5.3})^0)~~(2\sigma)&\nonumber\\
&\sin^2\theta_{12}=0.326^{+0.05}_{-0.04}~~~(\theta_{12}=(34.8^{+3.0}_{-2.5})^0)~~(2\sigma)&~~~.
\label{data}
\end{eqnarray}
\subsection{An orthodox viewpoint}
Despite all the recent experimental progress, the nature of $\delta{\cal L}(m_\nu)$ is still unknown. 
Perhaps, in the energy range presently explored, the best candidate for the new term is the leading non-renormalizable SU(2)$\times$U(1)
invariant operator.
To identify such operator, it is useful to think to particle interactions as described by a Lagrangian where the SM 
contribution, ${\cal L}_{SM}$, represents the first term of an expansion in inverse powers of a cut-off scale $\Lambda$:
 \begin{equation}
 {\cal L}={\cal L}_{SM}+\frac{1}{\Lambda} {\cal L}_5+\frac{1}{\Lambda^2} {\cal L}_6+...
 \label{weinberg}
 \end{equation}
While there are more than 80 independent dimension six operators in this expansion, there is a unique, up to flavor combination, dimension five operator
\cite{weinberg}:
\begin{equation}
\frac{{\cal L}_5}{\Lambda}=\frac{1}{\Lambda}({\tilde H}^\dagger l)^T Y  ({\tilde H}^\dagger l)+h.c.=\frac{v^2}{2\Lambda}\nu^T Y \nu+h.c.
\label{lfive}
\end{equation}
where $Y$ is a 3 $\times$ 3 complex symmetric matrix.
This operator provides neutrinos with a Majorana mass that is parametrically suppressed, with respect to the other fermion masses, 
by a factor $v/\Lambda$, $v\approx 246$ GeV being the electroweak scale. This allows us to interpret the smallness of neutrino masses in terms 
of the relative largeness of the cut-off scale $\Lambda$. A simple dimensional estimate suggests $\Lambda$ around $10^{15}$ GeV \cite{sha}, not very far from
the grand unified theory (GUT) scale, whose relevance is independently supported by gauge coupling unification and by considerations concerning matter stability.
If a new physical threshold around the GUT scale exists, neutrinos might offer a unique opportunity to explore an otherwise unaccessible
energy domain.
It is also a remarkable fact that, on the basis of the operator expansion in eq. (\ref{weinberg}), the first effect of new physics could have been
expected in the neutrino sector, where it was actually found! 

Several theoretical considerations support this picture. 
First of all  ${\cal L}_5$ violates $B-L$ by two units and such a violation is welcome.
Indeed all other global symmetries of the SM are violated and $B-L$ is violated
in many realistic GUTs. Moreover, global quantum numbers such as
$B-L$ are expected to be violated at least at the Planck scale by quantum gravity effects. 
The violation of $B-L$ is also crucial in baryogenesis, if the generation of the baryon asymmetry 
occurs at temperatures above the electroweak scale, where SM $B-L$ conserving sphaleron interactions
can erase any $B+L$ asymmetry.
\begin{figure}
\begin{center}
        \mbox{\epsfig{figure=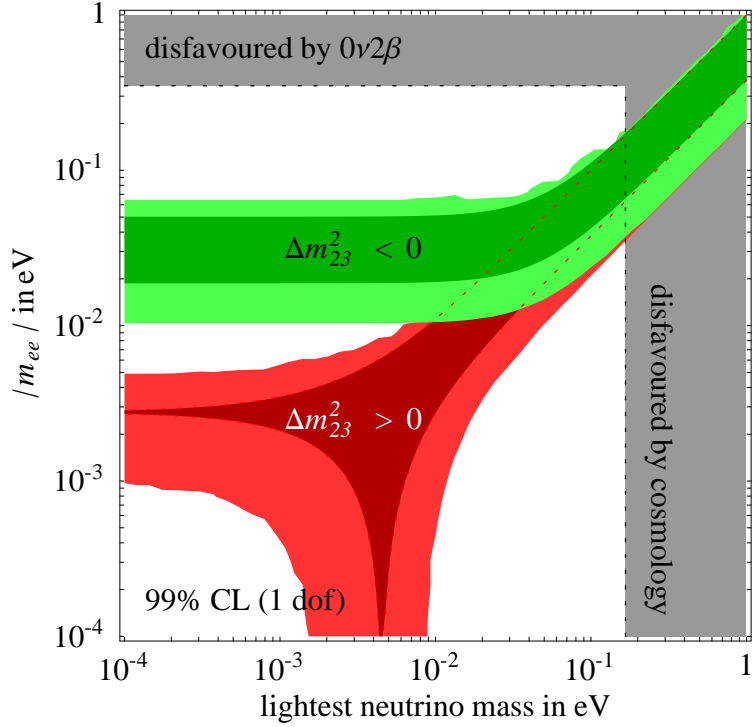,width=10.0cm}}
\end{center}
\caption{The parameter $|m_{ee}|$ of neutrinoless double beta decay, eq. (\ref{mee}), 
as a function of the lightest neutrino mass from ref. $^{7)}$.}
\label{0n2b}
\end{figure}
Many GUTs contain additional SM-singlet fermions, like the right-handed neutrinos $\nu^c$.
Heavy $\nu^c$ exchange produces a specific version of ${\cal L}_5$, through the see-saw mechanism \cite{seesaw}:
\begin{equation}
\frac{1}{\Lambda} {\cal L}_5=-\frac{1}{2}({\tilde H}^\dagger l)^T\left[y_\nu^T M^{-1} y_\nu\right] ({\tilde H}^\dagger l)+h.c.+...
\end{equation}
The see-saw mechanism can enhance small mixing angles in $M$ and $y_\nu$ into the large mixing angles
observed in neutrino oscillation. Moreover, the out-of-equilibrium, CP and $B-L$ violating
decay of right-handed neutrinos might offer a viable mechanism for generating the existing baryon asymmetry in the
universe \cite{leptogenesis}. Thus, from a theoretical viewpoint, $ {\cal L}_5/\Lambda$ is probably the best candidate
for $\delta {\cal L}(m_\nu)$, although a direct experimental evidence of it is lacking so far.  
\subsection{Tests of ${\cal L}_5/\Lambda$}
If $\delta {\cal L}(m_\nu)=  {\cal L}_5/\Lambda$, we expect neutrinoless double beta ($0\nu\beta\beta$) decay at some level. This process
is sensitive to the mass combination
\begin{equation}
\begin{array}{c}
\vert m_{ee} \vert=\vert\cos^2\theta_{13}(\cos^2\theta_{12} m_1+\sin^2\theta_{12} e^{2 i \alpha} m_2)
+\sin^2\theta_{13} e^{2 i \beta} m_3\vert
\end{array}
\label{mee}
\end{equation}
where $\alpha$ and $\beta$ are the two Majorana phases, not observable in neutrino oscillations. 
By using eq. (\ref{mee}) and by exploiting the present knowledge of neutrino masses and mixing angles, eqs. (\ref{data}), it is possible to predict the
range of $\vert m_{ee}\vert$ as a function of the lightest neutrino mass, see fig. 1.
A positive signal would allow to test $B-L$ violation and the absolute neutrino mass spectrum at the same time \cite{vogel}.
At present data set an upper bound at 90\% C.L. on $\vert m_{ee}\vert$ in the approximate range $0.2\div 1$ eV \cite{0n2b},
with a large uncertainty coming from poorly known nuclear matrix elements \cite{fogli,vogel}.  
There is also a claim of a positive signal in the same range from
the data of the Heidelberg-Moscow collaboration \cite{HM}. 
The expected sensitivity in $\vert m_{ee}\vert $ of the best forthcoming experiments
is close to  $0.01$ eV. Such a level would allow to completely test the case of inverted hierarchy in the neutrino mass spectrum. For an hierarchy of normal type unfortunately $\vert m_{ee}\vert$
can be much smaller than $0.01$ eV. 
Establishing the violation of $B-L$ and constraining the absolute neutrino mass spectrum  through $0\nu\beta\beta$ decay would represent a major step
in the field and any experimental effort aimed at improving the present sensitivity on $\vert m_{ee}\vert$ must be considered of the greatest importance.

\section{Low-energy tests of the neutrino mixing pattern}
Without any further assumption it is difficult to think of additional tests of the physics related to $\delta {\cal L}(m_\nu)$.
It is true that several important quantities such as the value of $\theta_{13}$, the neutrino mass ordering, the CP violating phases
and the absolute neutrino mass scale are still unknown. Their knowledge could add important restrictions on the fundamental theory
giving rise to $\delta {\cal L}(m_\nu)$.
Moreover, if the low-energy limit of the theory is correctly described by the expansion (\ref{weinberg}), we also expect
proton decay to occur. Its discovery and a detailed knowledge of its decay channels would provide
a wealth of important new constraints.
Nevertheless, even the simplest possible high-energy completion that can produce a viable $\delta {\cal L}(m_\nu)$, namely the (type I) see-saw mechanism
recalled above, is not easily testable. Indeed the theory that includes three right-handed neutrinos $\nu^c$ depends on 18 parameters in the
neutrino sector alone: six masses, six mixing angles and six phases, the double of those entering the low-energy description through the operator  
${\cal L}_5/\Lambda$.
\subsection{Tests of flavour symmetries}
A very interesting framework arises if the fundamental theory possesses some flavour symmetry, with a relevant scale of the order of
the cut-off $\Lambda$. Indeed this is suggested by the closeness of the observed lepton mixing to the so-called tri-bimaximal (TB) mixing pattern \cite{TB}:
\begin{eqnarray}
\sin^2\theta_{13}^{TB}&=&0\nonumber\\
\sin^2\theta_{23}^{TB}&=&1/2\nonumber\\
\sin^2\theta_{12}^{TB}&=&1/3~~~.
\label{thetatb}
\end{eqnarray} 
Such a pattern can be successfully reproduced in models
with spontaneously broken flavour symmetries and special vacuum alignment properties \cite{ma,us,others}.
In these models the flavour symmetry group $G_f$, which will be specified later on, 
is spontaneously broken by a set of adimensional small parameters $\langle\Phi\rangle$,
interpreted as ratios between vacuum expectation values (VEVs) of flavon scalar fields $\varphi$, transforming non-trivially under $G_f$,
and the cut-off scale $\Lambda$. The consistency of this picture requires that  $\vert\langle\Phi\rangle\vert\ll 1$.
In this scenario, the Yukawa couplings $y_f$ $(f=l,u,d)$ of the SM and the matrix $Y$ in ${\cal L}_5$
become functions of $\langle\Phi\rangle$:
\begin{equation}
y_f=y_f\left(\langle\Phi\rangle\right)~~~,~~~~~~~Y=Y\left(\langle\Phi\rangle\right)~~~,
\label{yuk}
\end{equation}
so that, by treating $\langle\Phi\rangle$ as spurions transforming under $G_f$ as the corresponding 
parent scalar fields, the whole theory is formally invariant under $G_f$.
Under the assumption that $\vert\langle\Phi\rangle\vert\ll 1$, the functions $y_f$ and $Y$ can be expanded in powers of
$\langle\Phi\rangle$ and only a limited number of terms gives a non-negligible contribution.

Even when $G_f$ and $\langle\Phi\rangle$ are completely specified, as in the case we are going to discuss,
if $\Lambda$ is much larger than the electroweak scale, we can test $G_f$ only at the level of the
fermion masses and mixing angles.
Instead it would be highly desirable to find evidence of the flavour symmetry in other types of processes.
Such a possibility becomes realistic if there is new physics at a much closer
energy scale $M$, around $1\div 10$ TeV. 
Indeed we have several indications, both from the experimental and from the theory point of view,
that this can be the case. For instance, the observed discrepancy in the anomalous muon magnetic moment, the overwhelming evidence 
of dark matter, the evolution of the gauge coupling constants towards a common high-energy value and the solution of the
hierarchy problem can all benefit from the existence of new particles around the TeV scale. 

If such a new scale exists, in an effective field theory approach the dominant physical effects of the new particles at low energies
can be described by dimension six operators, suppressed by two powers of the new mass scale $M$ and explicitly conserving $B$ and $L$,
if we assume that the new degrees of freedom do not provide new sources of baryon and/or lepton number violation.
If we focus on the lepton sector only, the leading terms of the relevant effective Lagrangian are:
\begin{equation}
{\cal L}_{eff}={\cal L}_{SM}+\delta {\cal L}(m_\nu)+
i\frac{e}{M^2} {e^c}^T H^\dagger \sigma^{\mu\nu} F_{\mu\nu} {\cal M}  l +h.c.+[\tt 4-fermion~~ operators]
\label{leff}
\end{equation}
where $e$ is the electric charge, $e^c$ the set of SU(2) lepton singlets and
$F_{\mu\nu}$ is the electromagnetic field strength
\footnote{Gauge invariance would allow $F_{\mu\nu}$ to be an arbitrary
combination of $B_{\mu\nu}$, the field strength of U(1), and $\sigma^a W^a_{\mu\nu}$, the non-abelian field strength of SU(2).}. 
The complex three by three matrix ${\cal M}$, with indices in the flavour space, is a function of $\langle\Phi\rangle$:
\begin{equation}
{\cal M}={\cal M}\left(\langle\Phi\rangle\right)~~~.
\end{equation}
The effective Lagrangian ${\cal L}_{eff}$ is invariant under $G_f$, once
we treat the symmetry breaking parameters as spurions. 
As a result, the same symmetry breaking parameters that control lepton masses and mixing angles also control
the flavour pattern of the other operators in ${\cal L}_{eff}$. Moreover the effects described by these operators
are suppressed by $1/M^2$ and not by inverse powers of the larger scale $\Lambda$ and this opens the possibility that
they might be observable in the future. 

These effects are well-known \cite{raidal}. In a field basis where the kinetic terms are canonical and the charged lepton
mass matrix is diagonal (we denote by a hat the relevant matrices in this basis), the real and imaginary parts of the matrix elements
$\hat{\cal M}_{ii}$ are proportional to the anomalous magnetic moments (MDM) $a_i$
and to the electric dipole moments (EDM) $d_i$ of charged leptons, respectively:
\begin{equation}
a_i=2 m_i \frac{v}{\sqrt{2} M^2}Re \hat{\cal M}_{ii}~~~,~~~~~~~d_i=e \frac{v}{\sqrt{2} M^2}Im \hat{\cal M}_{ii}~~~.
\label{dm}
\end{equation}
The off-diagonal elements 
$\hat{\cal M}_{ij}$ describe the amplitudes for the lepton flavour violating (LFV) transitions \cite{lfvvarie}
$\mu\to e \gamma$, $\tau\to\mu\gamma$ and $\tau\to e \gamma$:
\begin{equation}
R_{ij}=\frac{BR(l_i\to l_j\gamma)}{BR(l_i\to l_j\nu_i{\bar \nu_j})}=\frac{12\sqrt{2}\pi^3 \alpha}{G_F^3 m_i^2 M^4}\left(\vert\hat{\cal M}_{ij}\vert^2+\vert\hat{\cal M}_{ji}\vert^2\right)
\label{dt}
\end{equation}
where $\alpha$ is the fine structure constant, $G_F$ is the Fermi constant and $m_i$ is the mass of the lepton $l_i$.
Finally the four-fermion operators, together with the dipole
operators controlled by $\hat{\cal M}$, describe other flavour violating processes
like $\mu\to eee$, $\tau\to\mu\mu\mu$, $\tau\to eee$. 
\subsection{A special flavour symmetry: $G_f=A_4\times Z_3\times U(1)_{FN}$}
The framework described above finds its simplest realization in minimal flavour violation (MFV) \cite{MFV,MLFV},
where the flavour symmetry group $G_f$ contains, in the lepton sector, SU(3)$_{e^c}\times$ SU(3)$_l$, and where the dimensionless
symmetry breaking parameters are the Yukawa couplings themselves. While such a choice has the advantage that it can accommodate any
pattern of lepton masses and mixing angles, it does not provide any clue about the origin of the 
approximate TB pattern observed in the lepton mixing matrix $U_{PMNS}$.
For this reason, here we prefer to discuss the case
\begin{equation}
G_f=A_4\times Z_3\times U(1)_{FN}
\label{flavgroup}
\end{equation}
that has been especially tailored to approximately reproduce in a simple and economic way the TB mixing scheme \cite{us}.
The three factors in $G_f$ play different roles. The spontaneous breaking of the first one, $A_4$, is directly responsible for the TB mixing.   
The $Z_3$ factor is a discrete version of the total lepton number and is needed in order to avoid large mixing effects
between the flavons that give masses to the charged leptons and those giving masses to neutrinos. Finally, $U(1)_{FN}$ is
responsible for the hierarchy among charged fermion masses.
The flavour symmetry breaking sector of the model includes the scalar fields $\varphi_T$, $\varphi_S$, $\xi$ and $\theta$. The transformation properties of the lepton
fields $l$, $e^c$, $\mu^c$, $\tau^c$, of the electroweak scalar doublet $H$ and of the flavon fields have been recalled in table 1.
The following pattern of VEVs for the flavon fields
\begin{eqnarray}
\frac{\langle\varphi_T\rangle}{\Lambda}&=&(u,0,0)+O(u^2)\nonumber\\
\frac{\langle\varphi_S\rangle}{\Lambda}&=& c_b(u,u,u)+O(u^2)\nonumber\\
\frac{\langle\xi\rangle}{\Lambda}&=&c_a u+O(u^2)\nonumber\\
\frac{\langle\theta\rangle}{\Lambda}&=&t
\label{vevs}
\end{eqnarray}
where $u$ and $t$ are the small symmetry breaking parameters of the theory, guarantees that the lepton mixing
is approximately TB. 
It is possible to achieve this pattern of VEVs in a natural way, as the result of the minimization of the scalar potential \cite{us}.
\begin{table}[!ht] 
\centering
                \begin{math}
                \begin{array}{|c||c|c|c|c||c|c|c|c|c|}
                    \hline
                    &&&&&&&&& \\[-9pt]
                    \tt{Field} & l & e^c & \mu^c & \tau^c & H & \phit & \phis & \xi & \theta \\[10pt]
                    \hline
                    &&&&&&&&&\\[-9pt]
                    A_4 & 3 & 1 & \onepp & \onep & 1 & 3 & 3 & 1 &  1 \\[3pt]
                    \hline
                    &&&&&&&&&\\[-9pt]
                    Z_3 & \om & \om^2 & \om^2 & \om^2  & 1 & 1& \om & \om  & 1 \\[3pt]
                    \hline
                    &&&&&&&&&\\[-9pt]
                    U(1)_{FN} & 0 & 2 & 1 & 0  & 0 & 0 & 0 & 0 & -1  \\[3pt]
                    \hline
                \end{array}
               \end{math} 
            \caption{ The transformation rules of the fields under the
            symmetries associated with the groups \rm{$A_4$}, \rm{$Z_3$} and
            \rm{$U(1)_{FN}$}.}
            \end{table}
\vskip 0.2cm
The Yukawa couplings in eq. (\ref{yuk})
are given by:
\begin{equation}
y_l=\left(
\begin{array}{ccc}
y_e t^2 & 0& 0\\
0& y_\mu t& 0\\
0& 0& y_\tau
\end{array}
\right) u+O(u^2)~~~,
\label{yf}
\end{equation}
\begin{equation}
Y=\left(
\begin{array}{ccc}
a+2 b/3& -b/3& -b/3\\
-b/3& 2b/3& a-b/3\\
-b/3& a-b/3& 2 b/3
\end{array}
\right) u+O(u^2)~~~,
\label{Y}
\end{equation}
where $y_e$, $y_\mu$, $y_\tau$, $a$ and $b$ are numbers of order one ($a$ and $b$ are proportional to the coefficients $c_a$ and $c_b$ of eq. (\ref{vevs})).
At the leading order, neglecting the $O(u^2)$ contributions, the mass matrix for
the charged leptons is diagonal with the relative hierarchy described by the parameter $t$. 
To reproduce the correct hierarchy we will take
\begin{equation}
\vert t\vert\approx 0.05~~~.
\label{tbound}
\end{equation}
In the same approximation,
the neutrino mass matrix is diagonalized by the transformation:
\begin{equation}
U_{TB}^T m_\nu U_{TB} =\frac{v^2}{\Lambda}{\tt diag}(a+b,a,-a+b) u~~~,
\end{equation}
where $U_{TB}$ is the unitary matrix of the TB mixing:
\begin{equation}
U_{TB}=\left(
\begin{array}{ccc}
\sqrt{2/3}& 1/\sqrt{3}& 0\\
-1/\sqrt{6}& 1/\sqrt{3}& -1/\sqrt{2}\\
-1/\sqrt{6}& 1/\sqrt{3}& +1/\sqrt{2}
\end{array}
\right)~~~.
\label{UTB}
\end{equation}
The $O(u^2)$ contributions in eqs. (\ref{vevs}), (\ref{yf}) and (\ref{Y}) give rise
to corrections to the TB mixing of relative order $u$. 
The symmetry breaking parameter $u$ should lie in the range
\begin{equation}
0.001<\vert u\vert< 0.05~~~,
\label{ubound}
\end{equation}
the lower bound coming from the requirement that the Yukawa coupling of the $\tau$ does not exceed $4 \pi$, and the upper bound coming from the
requirement that the higher order corrections, so far neglected, do not modify too much the leading TB mixing.
Indeed, the inclusion of higher order corrections modifies all mixing angles by quantities of relative order $u$, 
especially we should keep the agreement between the
predicted and measured values of the solar angle within few degrees. The unknown angle $\theta_{13}$ is expected to be of order $u$,
not far from the future aimed for experimental sensitivity \cite{tonazzo}.
Such a framework can also be extended to the quark sector \cite{quarks}. Constraints from baryogenesis have been discussed in ref. \cite{mano}.
\vskip 0.5 cm
\subsection{Results for the dipole moments and for the lepton flavour transitions}
We can read off the observable quantities $a_i$, $d_i$ and $R_{ij}$ from eqs. (\ref{dm}) and (\ref{dt}) after evaluating the 
matrix of dipole moments $\hat{\cal M}$. Starting with the relevant set of invariant operators, after the breaking of the flavour and electroweak symmetries, 
and after moving to a basis with canonical kinetic terms and diagonal mass matrix for charged leptons, we find \cite{lfva4}:
\begin{equation}
\hat\mathcal{M} = \left( \begin{array}{ccc}
        O(t^2 u) & O(t^2 u^2) & O(t^2 u^2)\\
        O(t u^2) & O(t u) &  O(t u^2)\\
        O(u^2) & O(u^2) & O(u)
\end{array}
\right)
\label{hatM}
\end{equation}
where each matrix element is known only up to an unknown order one dimensionless coefficient.
We can see that MDMs and EDMs arise at the first order in the parameter $u$. By assuming that the unknown coefficients have absolute values and phases of order one, we have:
\begin{equation}
a_i=O\left(2\dd\frac{m_i^2}{M^2}\right)~~~,~~~~~~~~~~d_i=O\left(e\dd\frac{m_i}{M^2}\right)~~~.
\label{oom}
\end{equation}
We can derive a bound on the scale $M$, by considering the
existing limits on MDMs and EDMs and by using eqs. (\ref{oom}) as exact equalities to fix the ambiguity of the unknown coefficients.
We find the results shown in table 2.
\begin{table}[!ht] 
\centering
                \begin{math}
                \begin{array}{|c|c|}
                    \hline
                    & \\[-9pt]
                    d_e<1.6\times 10^{-27}~~e~cm&M>80~~{\rm TeV}\\[3pt]
                    \hline
                    &\\[-9pt]
                   d_\mu<2.8\times 10^{-19}~~e~cm&M>80~~{\rm GeV}\\[3pt]
                    \hline
                    &\\[-9pt]
                    \delta a_e<3.8\times 10^{-12}&M>350~~{\rm GeV}\\[3pt]
                    \hline
                    &\\[-9pt]
                    \delta a_\mu\approx 30\times 10^{-10}&M\approx 2.7~~{\rm TeV}\\[3pt]
                    \hline
                \end{array}
               \end{math} 
            \caption{Experimental limits on lepton MDMs and EDMs $^{23)}$ and corresponding bounds on the scale $M$, derived from eq. (\ref{oom}).
The data on the $\tau$ lepton have not been reported since they are much less constraining. For the anomalous magnetic moment of the muon,
$\delta a_\mu$ stands for the deviation of the experimental central value from the SM expectation $^{24)}$.}
            \end{table}
\vskip 0.2cm
\noindent
From table 2 we see that, in order to accept values of $M$ in the range $1\div 10$ TeV, we should invoke a cancellation in the imaginary part
of $\hat{\cal{M}}_{ee}$, which can be either accidental or due to CP conservation in the considered sector of the theory.
Concerning the flavour violating dipole transitions, from eq. ({\ref{hatM}) we see that the rate for $l_i\to l_j\gamma$ is dominated by
the contribution $\hat{\cal M}_{ij}$, since $\hat{\cal M}_{ji}$ is suppressed by a relative factor of $O(t)$ for $\mu\to e\gamma $ and $\tau\to\mu\gamma$
and of $O(t^2)$ for $\tau\to e\gamma$. We get:
\begin{equation}
\frac{BR(l_i\to l_j\gamma)}{BR(l_i\to l_j\nu_i{\bar \nu_j})}=\frac{48\pi^3 \alpha}{G_F^2 M^4}\vert w_{ij} ~u\vert^2
\label{LFV}
\end{equation}
where $w_{ij}$ are numbers of order one.
As a consequence, the branching ratios of the three
transitions $\mu\to e\gamma $, $\tau\to\mu\gamma$ and $\tau\to e\gamma$ should all be of the same order:
\begin{equation}
BR(\mu\to e \gamma)\approx BR(\tau\to\mu\gamma)\approx BR(\tau\to e \gamma)~~~.
\label{equalbr}
\end{equation}
This is a distinctive feature of this class of models, since in most of the
other existing models there is a substantial difference between the branching ratios. In particular it often occurs that
$BR(\mu\to e \gamma)<BR(\tau\to\mu\gamma)$ \cite{raidal}.  Given the present experimental bound $BR(\mu\to e \gamma)<1.2\times 10^{-11}$,
eq. (\ref{equalbr}) implies that $\tau\to\mu\gamma$ and $\tau\to e \gamma$ have rates much below the present and expected future sensitivity
\footnote{The present limits are $BR(\tau\to\mu\gamma)<1.6\times 10^{-8}$ and $BR(\tau\to e\gamma)<9.4\times 10^{-8}$. A future super B factory
might improve them by about one order of magnitude.}.
Moreover, from the current (future) experimental limit on $BR(\mu\to e \gamma)$ \cite{bemporad} and assuming $\vert w_{\mu e}\vert=1$, we derive the following bound
on $\vert u/M^2\vert$:
\begin{equation}
BR(\mu\to e \gamma)<1.2\times 10^{-11}~(10^{-13})~~~~~~~
\left\vert\dd\frac{u}{M^2}\right\vert<1.2\times 10^{-11}~(1.1\times 10^{-12})~~{\rm GeV}^{-2}~~~.
\end{equation}
\noindent
Since the parameter $\vert u\vert$ lies in the limited range of eq. (\ref{ubound}), we find
\begin{eqnarray}
\vert u\vert = 0.001 &~~~~~~~~~~M>10~(30)~~{\rm TeV}\\ 
\vert u\vert = 0.05 &~~~~~~~~~~~~~~M>70~(200)~~{\rm TeV}~~~.
\end{eqnarray}
This pushes the scale $M$ considerably above the range we were initially interested in. In particular $M$ is shifted above the region of 
interest for $(g-2)_\mu$ and for LHC.

\subsection{Supersymmetric case}
The off-diagonal elements of the dipole matrix ${\cal M}$ are all of order $u^2$ and they come from two independent sources.
They can originate either from operators with a single insertion of the flavon $\varphi_T$ through the $O(u^2)$ sub-leading corrections to its VEV,
or from operators where double flavon insertions are considered.
In this second case, the only combinations of flavon insertions that can provide a non-vanishing contribution are 
$\xi^\dagger \varphi_S$ and its conjugate. In a generic case we expect that both these terms are equally important
and contribute at the same order to a given off-diagonal dipole transition.
There is however a special case where the double flavon insertions of $\xi^\dagger \varphi_S$ and its conjugate
are suppressed compared to the sub-leading corrections to $\varphi_T$. This happens, under certain conditions, 
when the underlying theory is supersymmetric and supersymmetry is softly broken. 
We consider supersymmetry breaking originating in a hidden sector of the theory and transmitted to the observable sector
via gravitational interactions, so that soft supersymmetry breaking terms with a typical mass scale of order $m_{SUSY}$ 
are generated. We also assume that in the underlying fundamental theory the only sources of chirality flip are
either fermion masses or sfermion masses of left-right type. Both of them, up to the order $u^2$,
are dominated by the insertion of $\varphi_T$ or $\varphi_T^2$ in the relevant operators.
Insertions of non-holomorphic combinations, like $\xi^\dagger \varphi_S$, are suppressed by powers of $m_{SUSY}/\Lambda$.
This defines what we mean by ''supersymmetric case''.
 \begin{figure}[t]
\begin{center}
        \mbox{\epsfig{figure=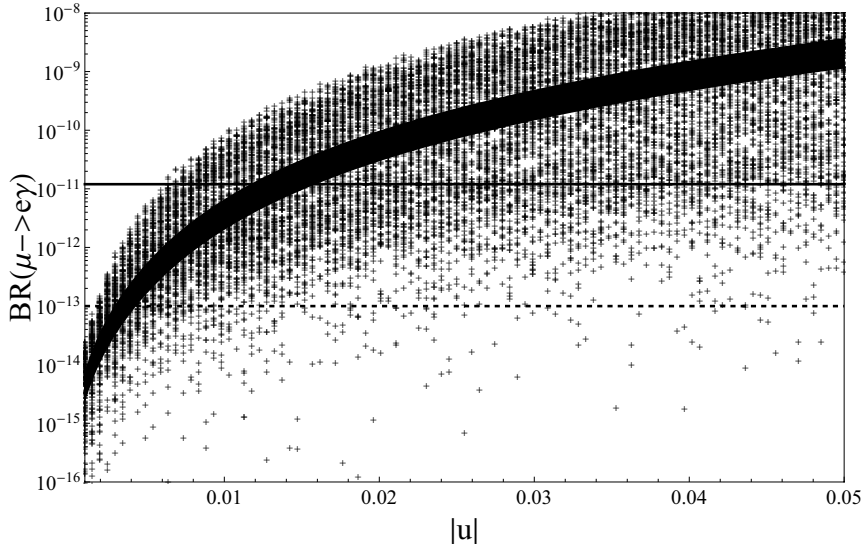,width=12.0cm}}
\end{center}
\caption{The branching ratio of $\mu\to e \gamma$ as a function of $|u|$, eq. (32). The deviation of the anomalous magnetic moment of the muon from the SM expectation is kept fixed to its experimental range. 
The unknown coefficients $\tilde{w}^{(1,2)}_{\mu e}$ are equal to 1 (darker region) or are random
complex numbers with absolute values between zero and two (lighter region).
The continuous (dashed) horizontal line corresponds to the present (future expected) experimental bound on $BR(\mu\to e\gamma)$.}
\label{BR}
\end{figure}
In the supersymmetric case we should take into account
the fact that any chirality flip up to the order $u^2$ necessarily requires the insertion of $\varphi_T$ or
$\varphi_T^2$. This restriction produces a cancellation in the elements below the diagonal of the matrix $\hat{\cal M}$.
Starting from the set of invariant operators allowed by these considerations and moving to a basis with canonical kinetic terms and 
diagonal mass matrix for charged leptons, we obtain \cite{lfva4}:
\begin{equation}
\hat\mathcal{M} = \left( \begin{array}{ccc}
        O(t^2 u) & O(t^2 u^2) & O(t^2 u^2)\\
        O(t u^3) & O(t u) &  O(t u^2)\\
        O(u^3) & O(u^3) & O(u)
\end{array}
\right)
\label{hatMsusy}
\end{equation}
The EDMs and MDMs are similar to those of the general non-supersymmetric case.
For the LFV transitions we get
\begin{equation}
\frac{BR(l_i\to l_j\gamma)}{BR(l_i\to l_j\nu_i{\bar \nu_j})}= \frac{48\pi^3 \alpha}{G_F^2 M^4}\left[\vert w^{(1)}_{ij} u^2\vert^2+\frac{m_j^2}{m_i^2} \vert w^{(2)}_{ij} u\vert^2\right]
\label{LFVsusy}
\end{equation}
where $w^{(1,2)}_{ij}$ are unknown, order one coefficients. 
Notice that now the contribution from $\hat{\cal M}_{ij}$ is suppressed by a factor of $u$ compared to the non-supersymmetric case.
In most of the allowed range of $u$, the branching ratios of $\mu\to e \gamma$ and $\tau \to\mu \gamma$ are similar and larger than
the branching ratio of $\tau\to e \gamma$.
Assuming $\vert w^{(1,2)}_{\mu e}\vert=1$,
the present (future) experimental limit on $BR(\mu\to e \gamma)$ implies the following bounds
\begin{eqnarray}
\vert u\vert = 0.001 &~~~~~~~~~~M>0.7~(2)~~{\rm TeV}\\ 
\vert u\vert = 0.05 &~~~~~~~~~~~~~~M>14~(48)~~{\rm TeV}~~~.
\end{eqnarray}
We see that at variance with the non-supersymmetric case there is a range of permitted values of the parameter $u$
for which the scale $M$ can be sufficiently small to allow an explanation of the observed discrepancy in $a_\mu$,
without conflicting with the present bound on $\mu\to e \gamma$. 
We can eliminate the dependence on the unknown scale $M$ by combining eqs. (\ref{oom}) and (\ref{LFVsusy}). For $\mu\to e \gamma$
we get:
\begin{equation}
\frac{BR(\mu\to e\gamma)}{BR(\mu\to e\nu_\mu{\bar \nu_e})}= \frac{12\pi^3 \alpha}{G_F^2 m_\mu^4}\left(\delta a_\mu\right)^2\left[\vert \tilde{w}^{(1)}_{\mu e}\vert^2 \vert u\vert^4+\frac{m_e^2}{m_\mu^2} \vert \tilde{w}^{(2)}_{\mu e}\vert^2\vert u\vert^2\right]
\label{muegamma}
\end{equation}
where $\tilde{w}^{(1,2)}_{\mu e}$ are unknown, order one coefficients. We plot $BR(\mu\to e\gamma)$ versus $|u|$ in fig. 2, where the coefficients $\tilde{w}^{(1,2)}_{\mu e}$ are kept fixed to 1 (darker region) or are random
complex numbers with absolute values between zero and two (lighter region). The deviation of the anomalous magnetic moment of the muon from the SM prediction is in the interval of the experimentally allowed values,
about three sigma away from zero. Even if the ignorance about the coefficients $\tilde{w}^{(1,2)}_{\mu e}$ does not allow us to derive a sharp limit on $|u|$, we see that the present limit on $\mu\to e \gamma$ disfavors values of $|u|$
larger than few percents. We recall that in this model the magnitudes of $|u|$ and $\theta_{13}$ are comparable. 

\subsection{Comparison with Minimal Flavour Violation}
It is instructive to compare the previous results with those of the MFV \cite{MLFV}.
If we restrict ourselves to the case where right-handed neutrinos do not affect
the flavour properties of the theory, the flavour symmetry group of MFV is $G_f=SU(3)_{e^c}\times SU(3)_l\times ...$,
where we have only displayed the part relevant to the lepton sector. Electroweak singlets $e^c$ and doublets $l$ transform as 
$(3,1)$ and $(1,\bar{3})$, respectively. The flavon fields or, better, their VEVs are the Yukawa couplings $y_l$ and $Y$ themselves,
transforming as $(\bar{3},3)$ and $(1,6)$, respectively. By going to a basis where the charged leptons are diagonal, $y_l$ and $Y$
can be expressed in terms of lepton masses and mixing angles (we keep using a notation where a hat denotes matrices in this particular
basis):
\begin{equation}
\hat{y}_l=\frac{\sqrt{2}}{v} m_l^{\rm diag}~~~,~~~~~~~~~~\hat{Y}=\frac{\Lambda}{v^2} U^* m_\nu^{\rm diag} U^\dagger~~~,
\end{equation} 
where $U$ is the lepton mixing matrix.
The diagonal elements of the matrix $\hat{\cal M}$ evaluated in MFV are analogous to those of the previous class of models and similar bounds
on the scale $M$ are derived from the existing data on MDMs and EDMs. The off-diagonal elements are given by:
\begin{eqnarray}
\hat{\cal M}_{ij}&=&\beta (\hat{y}_l \hat{Y}^\dagger \hat{Y})_{ij}\nonumber\\
&=&\sqrt{2}\beta\frac{m_i}{v}\frac{\Lambda^2}{v^4}\left[\Delta m^2_{sol} U_{i2} U^*_{j2}\pm\Delta m^2_{atm} U_{i3} U^*_{j3}\right]
\end{eqnarray}
where $\beta$ is an overall coefficient of order one and the plus (minus) sign refers to the case of normal (inverted) hierarchy.
We see that, due to the presence of the ratio $\Lambda^2/v^2$ the overall scale of these matrix elements is much less constrained than in the previous case.
This is due to the fact that MFV does not restrict the overall strength of the coupling constants $Y$, apart from the requirement 
that they remain in the perturbative regime.
Very small or relatively large (but smaller than one) $Y$ can be accommodated by adjusting the scale $\Lambda$.
On the contrary this is not allowed in the case previously discussed where the size of the symmetry breaking effects is restricted to the small window (\ref{ubound})
and the scale $\Lambda$ is determined within a factor of about fifty. 
The conclusion is that in MFV the non-observation of $l_i\to l_j \gamma$ could be justified by choosing a small $\Lambda$, while a positive signal
in $\mu\to e \gamma$ with a branching ratio in the range $1.2\times 10^{-11}\div 10^{-13}$ could also be fitted by an appropriate $\Lambda$,
apart from a small region of the $\theta_{13}$ angle, around $\theta_{13}\approx0.02$ where a cancellation in $\hat{\cal M}_{\mu e}$ can take place.

The dependence on the scale $\Lambda$ is eliminated by considering ratios of branching ratios. For instance:
\begin{equation}
\frac{BR(\mu\to e\gamma)}{BR(\mu\to e\nu_\mu{\bar \nu_e})}\frac{BR(\tau\to \mu\nu_\tau{\bar \nu_\mu})}{BR(\tau\to \mu\gamma)}=
\left\vert\frac{2\Delta m^2_{sol}}{3\Delta m^2_{atm}}\pm \sqrt{2}\sin\theta_{13} e^{i\delta}\right\vert^2<1~~~,
\label{mfv}
\end{equation}
where we took the TB ansatz, eq. (\ref{thetatb}), to fix $\theta_{12}$ and $\theta_{23}$. 
We see that $BR(\mu\to e\gamma)<BR(\tau\to \mu\gamma)$ always in MFV. Moreover, for $\theta_{13}$ above approximately $0.07$, $BR(\mu\to e\gamma)<1.2\times 10^{-11}$ implies $BR(\tau\to \mu\gamma)<10^{-9}$. For $\theta_{13}$ below $0.07$, apart possibly from a small region around $\theta_{13}\approx0.02$, both the transitions $\mu\to e \gamma$ and $\tau\to\mu\gamma$
might be above the sensitivity of the future experiments.
\section{Conclusion}
It should be clear from the above considerations that a theory of neutrino masses does not exist today.
We are still lacking a unifying principle that organizes the flavour sector of particle physics.
In a sense the current situation resembles that of electroweak interactions before the advent of
the Glashow-Weinberg-Salam electroweak theory. In the SM the pointlike Fermi interactions have been 
replaced by universal interactions with heavy gauge vector bosons, completely determined
by the principle of gauge invariance in terms of only two gauge coupling constants.
So far there is nothing comparable to the principle of gauge invariance in the flavour sector, where the interactions between
fermions and spin zero particles require many free parameters and where our ignorance of the neutrino sector
is probably best encoded in non-renormalizable interactions. We have several hints coming
from data and they have supported ideas and prejudices such as the existence of an underlying,
still to be discovered, flavour symmetry. 
Our hope is that new developments will be able to 
suggest how the flavour symmetry really acts, what is its symmetry breaking sector, what are the 
observable consequences and what is the interplay with more fundamental constructions
like GUTs or string theory. We are of course aware that, especially in the neutrino field and
in the recent past, many ideas and prejudices turned out to be wrong.

\section{Acknowledgements}
We thank the organizers of 
''{\it Nu HoRIzons}'' (February 13-15, 2008, Allahabad, India), 
''{\it Neutrino Oscillations in Venice}'' (April 15-18, 2008, Venice, Italy),
''{\it Melbourne Neutrino Theory Workshop}'' (June 2-4, 2008, Melbourne, Australia) and 
''{\it International School of Astroparticle Physics}'' (July 16-26, 2008, Valencia, Spain)
for giving us the opportunity to report on our recent work.
This work has been partly supported by the European Commission 
under contracts MRTN-CT-2004-503369 and MRTN-CT-2006-035505
and by the ``Sonderforschungsbereich'' TR27.

\end{document}